# A Layered Spin 1/2 polymorph of titanium triiodide

Danrui Ni,[a] Ranuri S. Dissanayaka Mudiyanselage,[b] Xianghan Xu,[a] Junsik Mun,[d] Yimei Zhu,[d] Weiwei Xie,[c] and Robert J. Cava[a]

[a]Department of Chemistry, Princeton University, Princeton, NJ 08544, USA

[b]Department of Chemistry and Chemical Biology, Rutgers University, Piscataway, NJ, 08854, USA

[c]Department of Chemistry, Michigan State University, 578 S Shaw Lane, East Lansing, MI 48824, USA

[d]Condensed Matter Physics and Materials Science Department, Brookhaven National Laboratory, Upton, New York 11973, USA

**Abstract**

A previously unreported layered spin ½ triangular lattice polymorph of $TiI_3$ is described, synthesized under 6 GPa of applied pressure at 900 °C, but stable at atmospheric pressure. This air-sensitive material has a $CdI_2$-type layered structure ($P$-$3m1$ (#164), $a$ = 4.012 Å and $c$ = 6.641 Å at 120 K, $Z$ = 1 of $Ti_{0.667}I_2$) with an in-plane triangular lattice, related to that of $TiI_4$ ($Ti_{0.5}I_2$). Although the $TiI_3$ formula is consistent with expectations for a layered honeycomb lattice of spin ½ Ti(III), there is disorder in the crystal structure. Magnetic susceptibility and heat capacity measurements suggest that the material undergoes several low temperature phase transitions.

*Keywords:* Titanium triiodide; High pressure synthesis; Spin ½; Layered structure; Polymorphism.

**Introduction**

Metal halides, especially transition metal halides, have attracted much research interest in recent years. Many members in this large family including 3D perovskites [1], 2D layers (triangular [2,3], Kagomé [4], or honeycomb [5–7]), and 1D chains [8,9] have been reported. Polymorphic behavior is also known [10–12]. The materials can display unexpected physical and/or photochemical properties, and thus promising performance in photovoltaic, electric, and quantum material applications [13–16]. It is therefore of great interest to characterize new phases of transition metal halides and study their physical and chemical properties. Here we describe the disordered layered triangular lattice spin ½ material $TiI_3$ synthesized at 6 GPa, its crystal structure is very different from that of the ambient pressure synthesized phase, which is made from titanium iodide chains.

Among the reported transition metal trihalides, TiCl$_3$, with Ti$^{3+}$ in $d^1$ configuration, was found to have both layered and chain-structure polymorphs as early as 1961 [10]. The triiodide variant, TiI$_3$, however, has only been reported to display a 1D chain structure. Iodides can display properties that are much different from their chloride cousins and are therefore worth finding [17], and a potential two-dimensional phase consisting of spin ½ metals in a layered lattice with iodine atoms is theoretically expected to display interesting electronic and magnetic properties [18]. The layered polymorph of spin 1/2 TiI$_3$ has not been reported experimentally until now. We report it here but find it to display a disordered Ti sublattice when synthesized under our conditions, which may be due to either stacking faults or in-plane Ti-vacancy disorder, the latter being the most likely.

As a starting point for understanding the synthesis of the current material, we note that pressure-induced polymorphism has been reported in the well-studied ruthenium trihalides. While RuCl$_3$ undergoes a phase transition from its chain-structure β-phase to a honeycomb-layered α-phase with increasing temperature, for its sister compounds RuBr$_3$ and RuI$_3$ only chain structures have been reported at ambient pressure; with the help of applied pressure, however, the honeycomb layer polymorphs of RuBr$_3$ and RuI$_3$ have been experimentally synthesized [17,19,20]. In that vein, here we employ modest temperatures and pressures to synthesize a layered polymorph of spin ½ TiI$_3$. Layered TiI$_3$ is air-sensitive (as is the ambient pressure polymorph) and is refined to have a *P*-3*m*1 CdI$_2$-type structure with $d^1$ Ti(III) on average partially occupying 2/3 of the Cd-type sites. This spin ½ material is structurally related to the $d^0$ compound TiI$_4$ (Ti$_{0.5}$I$_2$) whose structure consists of zig-zag chains of edge-shared TiI$_6$ octahedra within a layered iodine array. Possible explanations for the partial occupancy of titanium in the triangular layered lattice are discussed, and magnetic susceptibility and heat capacity measurements are carried out, suggesting the presence of phase transitions in TiI$_3$ in the lower temperature range.

**Experimental**

The ambient-pressure phase of TiI$_3$, used as a starting material, was synthesized using elemental titanium powder (Alfa Aesar, 99.9%) and iodine (Sigma Aldrich, 99.99%). It was annealed in vacuum in sealed quartz tubes for three days, with multiple annealing temperatures between 300 – 700 °C tested to increase the yield. Some impurities were produced during the annealing, including TiI$_4$, and TiI$_2$. A low temperature vapor transport method was used for purification before the high-pressure synthesis. With the hot zone kept at 350 °C, impurities such as TiI$_4$ are transferred to the cold end, leaving pure TiI$_3$ in the region with higher temperature. The resulting black-colored, 1D-chain structure TiI$_3$ material was used as starting material for the high-pressure synthesis. TiI$_4$ was also employed for high pressure synthesis experiments, as described below.

The ground powder of the starting material was loaded into a boron nitride crucible and then inserted into a pyrophyllite cube assembly. The samples were pressed to 6 GPa using a cubic multi-anvil system (Rockland Research Corporation) and annealed at different temperatures (700 – 1000 °C), with the temperature measured by an internal thermocouple. The samples were then quench-cooled to ambient temperature before decompression. Annealing at 900 °C for 1 hour gave the best results. The products obtained displayed a black color. Small single crystals were picked up in the post-reaction samples and were used for single crystal X-ray diffraction (SCXRD) characterization of the crystal structure. All the titanium iodide compounds prepared require handling in an air-free atmosphere. Thus the PXRD patterns used for characterization were collected using a Rigaku Miniflex II diffractometer located inside a nitrogen-filled glove box. Cu Kα radiation (λ= 1.5406 Å) was employed, and Le Bail fitting of the acquired patterns, when performed, was conducted via the TOPAS software.

SCXRD data of $TiI_3$ crystals was collected at 120 K on a Bruker D8 Quest Eco using graphite-monochromated Mo Kα radiation (λ = 0.71073 Å). A liquid nitrogen stream was used to prevent samples from decomposing. The frames were integrated using the SAINT program within the APEX III 2017.3-0 operating system. The structure was determined using direct methods and difference Fourier synthesis (SHELXTL version 6.14) [21]. The *P*-3*m*1 (#164) space group was suggested by XPREP. Other potential space groups (*P*-31*c*, *P*63*mc*, *P*63/*mmc* and *P*6/*mmm*) were also tested but structures in those space groups did not lead to satisfactory or better refinements, with either higher R/wR2 or higher $R_{int}$ values. The electron microscopy measurements were conducted on ground powder of $TiI_3$ but the beam-sensitivity of the sample significantly limited its characterization by this method.

The extreme air sensitivity of the high-pressure-synthesized compound makes getting quantitative magnetic susceptibility and heat capacity data difficult, but magnetization and heat capacity measurements were conducted in any case, with an effort made to minimize contact of the material with air. The experiments were carried out using a Quantum Design PPMS (Dynacool), equipped with a vibrating sample magnetometer (VSM) option. For these and all other cases, transfers were performed very rapidly to prevent sample decomposition. Magnetic susceptibility was defined as M/H, and the temperature-dependent magnetization (M) was measured in an applied magnetic field (H) of 1000 Oe.

**Results and Discussion**

1D chain $TiI_3$ (*Pmnm* [22]) and cubic $TiI_4$ (*Pa*-3 [23]) were synthesized and purified at ambient pressure, and were used as starting materials for the high-pressure synthesis (Figures S1 and S2 show their ambient temperature PXRD patterns). Optimized synthesis conditions for the high pressure $TiI_3$ phase were found to be 900 °C and 6 GPa for 1 hour. The high-pressure $TiI_3$ phase (HP-$TiI_3$) obtained crystallizes in a layered structure in the *P*-3*m*1 space group (#164) with *a* =

4.012 Å and *c* = 6.641 Å at 120 K (Figure 1). The crystallographic information is presented in Tables 1-3 and Table S1. During the refinement, pseudo symmetry elements (i.e. a $6_3$ screw and a *c* glide) were suggested by Checkcif but they were found to not be present. The *P*-3*m*1 structure yields the best refinement. The hk0, hk1 and 0kl reciprocal planes of HP-TiI$_3$ are shown in Figure 2A.

HP-TiI$_3$ crystallizes in a CdI$_2$-type layered structure with a triangular lattice, where the Ti on average randomly occupies the Cd site, leading to an occupancy ratio of 0.667Ti:2I when considered on an MX$_2$ structural basis. Compared to the low-temperature structure of ambient pressure chained TiI$_3$, which was reported at approximately the same temperature [22], the TiI$_6$ octahedra in the HP material are less distorted. The Ti-I bond length is about 2.82 Å, and the TiI$_6$ polyhedron is closer to that of a regular octahedron. When compared to high-temperature chain-structure AP-TiI$_3$ (*P*63/*mcm* space group with more symmetric Ti-coordination), the bonding distance is slightly longer in HP-TiI$_3$ (Table 3). The 300 K PXRD pattern of HP-TiI$_3$ is consistent with the SCXRD results (confirmed by Le Bail fitting in Figure S3), yielding an *a* of 4.0277(2) Å and a *c* of 6.6828(2) Å at 300 K for the *P*-3*m*1 unit cell. A small amount of ambient pressure TiI$_3$ is found in the HP-TiI$_3$ PXRD pattern - less than 9% based on a rough Rietveld refinement.

After an analogous high-pressure high-temperature treatment, TiI$_4$ shows a PXRD pattern that indicates that it also has a trigonal crystal structure. Some impurities remain present, including monoclinic TiI$_4$, whose crystal structure [24] is closely related to that of trigonal HP-TiI$_3$. 1D chain TiI$_3$ and titanium metal are also present (Figure S4). The cubic TiI$_4$ polymorph, which was the starting material for the HP synthesis, is reported to be metastable at 300 K, slowly transforming to monoclinic TiI$_4$ at ambient pressure [23], so it is not surprising that this phase was not observed in the post-reaction PXRD pattern. By viewing the PXRD patterns of the high pressure synthesized TiI$_3$ and TiI$_4$ together (Figure 2B), small shifts of the peaks are revealed. The similar diffraction patterns suggest that HP-TiI$_4$ and HP-TiI$_3$ are essentially isostructural, but that due to the different Ti:I ratios in the starting materials, HP-TiI$_4$ has a different occupancy of Ti on the 1a Wyckoff site in the crystal structure. The Le Bail fitting of the 300 K PXRD pattern for HP-TiI$_4$ yields *a* = 4.0178(2) Å and *c* = 6.6299(3) Å, with these smaller values being consistent with the expected lower radius of $d^0$ Ti$^{4+}$ compared to that of $d^1$ Ti$^{3+}$. This structure type has also been observed for TeI$_4$ [25], as one of its polymorphs was found to adopt a CdI$_2$-type structure (Te$_{0.5}$I$_2$) with a random distribution of Te and vacancies over the potential metal positions in the layered triangular lattice. Another comparable example is Os$_x$Cl$_3$ [26,27], which will be discussed in a later section. Based on the similarity of the two PXRD patterns, the HP titanium iodide series in the composition range studied may be a solid solution with the same *P*-3*m*1 layered structure but with different Ti content on the 1a site, i.e. resulting in an average formula of Ti$_x$I$_2$, with *x* varying at least between 0.5 and 0.667. As some reports suggest that TiI$_2$ also crystallizes in a $CdI_2$-type structure [28], the range of possible *x* values may be even larger, 0.5 to 1.

Thus $TiI_3$ undergoes a phase-transition from a 1D chain structure (Figure 1B) to a layered one at high pressures and temperatures. Similar polymorphism has been reported for $RuBr_3$ and $RuI_3$ [17,19,20]. The layered polymorphs of these compounds usually adopt a honeycomb Ru sublattice, however. As shown in Figure 3A and 3B, while a fully occupied triangular lattice leads to a 1:2 cation-to-anion ratio (i.e. a $CdI_2$ type structure), a honeycomb arrangement is an ordered 2/3 filled triangular lattice, resulting in a 0.667:2 = 1:3 formula ($BiI_3$ type), and a crystallographic cell in the basal plane that is about $\sqrt{3}$ larger than that observed here. We see no indication of this supercell in our diffraction patterns - in $TiI_3$ synthesized by our method, in contrast, it appears that the Ti atoms are on average randomly distributed on the metal (1a) sites of the triangular lattice. This may either be a truly random arrangement of Ti in the plane, or it may be that the local structure is still honeycomb-like in the plane with errors in the plane-stacking, making the average structure appear to be trigonal with a smaller in-plane cell. (SCXRD measurements are a positional average over the whole crystal, and thus a random layer stacking can give rise to a triangular reciprocal lattice in plane, with an equal occupancy fraction on each site of $Ti_{0.667}$). This explanation can also work for the HP-$TiI_4$ case. The thermodynamically stable phase of $TiI_4$ at ambient pressure has been refined to have *C*2/*c* symmetry [24], and when ignoring the slight distortion of $TiI_6$ octahedra, this structure can be considered as stacked $Ti_{0.5}I_2$ planes, with zig-zag chains of edge sharing $TiI_6$ octahedra in the basal-plane alternating with empty octahedra of similar geometry, as shown in Figure 3C. These intralayer zig-zag chains can be regarded as a 1/2 filled triangular lattice, matching the 0.5:2 = 1:4 formula. After annealing at high temperature and high pressure, the structure of $TiI_4$ appears to be less distorted, with more stochastic stacking of the planes, which are now more regularly stacked. Thus the PXRD pattern of HP-$TiI_4$ reflects an average structure close to that HP-$TiI_3$, but with different Ti fraction and different lattice parameters. Based on the SCXRD characterization of HP-$TiI_3$, however, evidence for the presence of interlayer stacking faults (streaks can be observed in 0kl reciprocal plane in Figure 2A, compared to the hk0 and hk1 planes) is relatively weak, and different from the expectation that there should be stronger streaking along l for interlayer stacking errors (also see the electron microscopy result shown in Figure S5). This observation makes in-plane disorder more likely. Characterization of the potential stacking faults, short-range in-plane Ti-vacancy arrangements, and the symmetry of smaller domains in both $TiI_3$ and $TiI_4$ may therefore be of future interest.

The resistance of dense polycrystalline pellets of HP-$TiI_3$ was too high to measure at room temperature, indicating that the material is either an insulator or a semiconductor. The magnetic behavior and thermal properties of spin ½ layered HP-$TiI_3$ were also estimated. The magnetic susceptibility measured on a polycrystalline HP-$TiI_3$ powder sample from 1.8 to 300 K is shown in Figure 4A (main panel), with data for 1D-chain structure ambient pressure AP-TiI$_3$ and HP-TiI$_4$ plotted in the same panel for comparison. In contrast to the other two materials, the behavior of HP-TiI$_3$ displays several features in the magnetic susceptibility below about 100 K, including a weak broad bump around 90 K, a feature around 45 K, and a transition at 22 K. These features are not observed in the χ vs. T plots for either AP-TiI$_3$ or HP-TiI$_4$. This indicates that these magnetic properties of HP-TiI$_3$ may be correlated with its layered structure and the spin ½ character of Ti(III). The structural disorder may introduce complexity in the magnetic behavior, however. Both zero-field-cooled (ZFC) and field-cooled (FC) data were collected on HP-TiI$_3$, and an obvious difference in the magnetic susceptibility measured in the ZFC and FC runs can be observed at temperatures below 20 K (Figure S6).

Magnetization versus applied magnetic field data for HP-TiI$_3$ were collected at different temperatures; the curves at 2 K, 15 K, 22 K, 35 K, and 75 K are presented in Figure 4B. An S-shape variation of M vs. H, with hysteresis, can be clearly observed at 2 K, with a magnetization that saturates to values slightly above 0.1 µ$_B$ per Ti (consistent with a canted-spin structure). The S-shape is also seen in the data taken at 15-35 K (Figure 4B inset), shrinking with increasing temperature and becoming a straight line in the 75 K data. This suggests that there may be a ferromagnetic component or some spin freezing or spin-glassiness component to the transition seen in the magnetic susceptibility. An X-ray diffraction study of a sample removed from the magnetometer under normal conditions showed the presence of some titanium (IV) iodide hydroxide hydrate (TiI(OH)$_3$·xH$_2$O) and iodine oxides in addition to the original HP-TiI$_3$ phase, as a result of the partial decomposition of the extremely air-sensitive material, and thus though the measured magnetic behavior can be considered as correct in general, we cannot guarantee that it is correct in detail.

We can speculate that the unusual magnetic behavior of spin-1/2 HP-TiI$_3$ may be due to multiple possible factors, including the spin ½ nature of Ti(III), structural disordering of the metal cations and vacancies, and the potential local distortion of the Ti(III)I$_6$ octahedra at low temperature, especially as symmetry breaking and dimerization of Ti(III) cations has been observed and reported for chain structured AP-TiI$_3$. It has been suggested that the distortion and breaking of intralayer trigonal symmetry can lead to significant effects on the electric and magnetic behavior of two-dimensional halides [29], and therefore it may be of future interest to study any potential transitions and low-temperature in-plane re-arrangements of the spin ½ Ti(III) ions in HP-TiI$_3$, as well as their correlation with magnetic properties.

The heat capacity of several HP-TiI$_3$ samples was measured from 1.8 to 120 K, without any magnetic field applied (Figure 4C). To roughly estimate the magnetic entropy, a fitting of the phonon contribution to the measured heat capacity using the Debye equation (Eq. (1)) was carried out. Considering the large atomic mass difference between Titanium and Iodine, a modified Debye equation:

$$C_{phonon} = 9R \sum_{n=1}^{2} C_n \left(\frac{T}{\Theta_{Dn}}\right)^3 \int_0^{\Theta_{Dn}/T} \frac{x^4 e^x}{(e^x - 1)^2} dx$$

(1)

was used, where $C_1$ = 1, and refined values of $\Theta_{D1}$ = 420.(6) K for titanium, and $C_2$ = 3, with a refined value of $\Theta_{D2}$ = 148.(7) K for iodine. (A similar modified Debye equation has been applied to fit the phonon heat capacity of other solids with coexisting light and heavy atoms [30,31].) The fitting was applied to the data between 40-60 K, which is consistent with the data in the high temperature range, as shown as the red curve. After subtracting this non-magnetic contribution from the heat capacity data (Figure 4C inset), a broad peak from 1.8 to 45 K is revealed, which appears to correspond to the magnetic features seen in the magnetic susceptibility data. The broad shape of the estimated $C_{mag}$ transition, as well as the fact that no sharp anomalies are observed in heat capacity data may be an indication of no long-range magnetic ordering, and that spin-freezing behavior, typical of spin glasses, is observed in this system. The glassiness may be due to either in-plane disorder or magnetic frustration, as the geometrical frustration that results from the triangular lattice as well as the mixture of disordered $Ti^{3+}$ and vacancies may suppress the long-range magnetic order. An estimated entropy change of 5.65 J/mol/K is calculated based on this transition. This number is very close to the value expected for Heisenberg spins (Rln(2S+1)) for S = ½ or Ising spins (Rln2), and thus is consistent with the expected electron configuration of Ti(III). An additional peak in the heat capacity is observed at around 85 K, which may correspond to a structural transition. It is known that the honeycomb layered polymorphs of the sister compounds $TiCl_3$ and $TiBr_3$ ($\alpha$-$TiCl_3$/$\alpha$-$TiBr_3$), undergo lattice distortions and dimerization at low temperature [32–34]. Thus transitions are also possible for the layered triiodide, although with a different in-plane lattice it may display different structural and magnetic properties.

Layered $Os_xCl_3$ (x = 0.825 [26] or 0.81 [27]) is a similar system, with disordered transition metal positions in a triangular in-plane lattice, resulting in an $CdI_2$-type average structure. Structural studies on that material reveal the possibility of short- and long-range vacancy ordering, and further suggest that different in-plane lattice arrangements may be achieved by different synthesis conditions, with an effect on the magnetism. Both $Os_xCl_3$ studies reported the suppression of long-range antiferromagnetic order and small, broad, field-dependent anomalies in the heat capacity, which is comparable to what is reported here for HP-$TiI_3$, reinforcing our observations

**Conclusions**

A layered-structure polymorph of titanium triiodide, which has not been reported previously, has successfully been made from the ambient pressure phase by high-pressure, high-temperature annealing. Different from the 1D chain structure reported for ambient pressure $TiI_3$, its structure is determined by SCXRD to be $CdI_2$-type, with a P-3m1 (#164) space group and unit cell ($a$ = 4.012 Å and $c$ = 6.641 Å at 120 K). The Ti site, with 66.7% occupancy on an ideal triangular lattice, may appear to be randomly occupied either be due to random stacking of the layers or to in-plane disorder. The magnetic or structural evidence for either of those things is weak at this point, however, and the spin ½ character of Ti(III) makes the local arrangement of the Ti an important question to address in future work. Our HP-TiI$_4$ samples present a similar structure after the same high-pressure treatment, with a different fraction of metal on the Ti site. An extensive layered triangular solid solution of M$_x$I$_2$ is proposed. Magnetic susceptibility and heat capacity measurements conducted on HP-TiI$_3$ reveal some magnetic features below 45 K and a large ZFC-FC deviation below 20 K. The polymorphism, structural properties, and unusual thermal and magnetic behavior of this disordered spin ½ material, which may act as a quantum material at low temperatures due to the low spin involved, may be of interest for further study, although it's air-sensitivity may inhibit its characterization by many techniques.

**Acknowledgements**

This research was funded in large part by the Gordon and Betty Moore foundation, EPiQS initiative, grant GBMF-9066. The single crystal diffraction work by Weiwei Xie's group was supported by U.S. DOE-BES under Contract DE SC0022156. The electron microscopy work at BNL was supported by DOE/BES, Division of Materials Science and Engineering under Contract No. DE-SC0012704.

**Appendix**

CCDC deposition number for HP-TiI$_3$: 2209900.

**Table 1.** Single crystal structure refinement for HP-TiI$_3$. (Standard deviation is indicated by the values in parentheses)

| Empirical Formula | Ti$_{0.67}$I$_2$ |
|---|---|
| Temperature (K) | 120(2) |
| Crystal System | Trigonal |
| Space Group | *P-3m1*; (#164) |
| Z | 1 |
| F.W. (g/mol) | 285.89 |
| *a*(Å) | 4.0118(6) |
| *c*(Å) | 6.6409(13) |
| V (Å$^3$) | 92.56(3) |
| F(000) | 121 |
| θ range (º) | 3.067 – 26.367 |
| Number of Reflections | 416; [$R_{int}$ = 0.0161] |
| Independent Reflections | 95 |
| Number of Parameters | 6 |
| Goodness-of-fit on F$^2$ | 1.368 |
| Final R Indices [I > σ(I)] | R1 = 0.0316, wR2 = 0.0773 |
| R Indices (all data) | R1 = 0.0318, wR2 = 0.0774 |
| Largest Diff. Peak and Hole (e$^-$/ Å$^3$) | 2.911; -1.420 |

**Table 2.** Atomic coordinates and equivalent isotropic displacement parameters of HP-TiI$_3$ at 120(2) K. (U$_{eq}$ is defined as one-third of the trace of the orthogonalized U$_{ij}$ tensor (Å$^2$)).

| Atom | Wyckoff. | Occ. | x | y | z | U$_{eq}$ |
|---|---|---|---|---|---|---|
| Ti1 | 1a | 0.6667 | 0 | 0 | 0 | 0.0166(19) |
| I1 | 2d | 1 | 0.666667 | 0.333333 | 0.75715(19) | 0.0125(5) |

**Table 3.** Selected bond lengths and bond angles for HP-TiI$_3$ (this work, refined at 120(2) K) and 1D-chain structure AP-TiI$_3$; low-temperature phase (LT) reported at 100 K and high-temperature phase (HT) reported at 326 K [22]. (Standard deviation is indicated by the values in parentheses)

|  | **HP-TiI$_3$** |  | **AP-TiI$_3$ (LT)** [22] |  | **AP-TiI$_3$ (HT)** [22] |  |
|---|---|---|---|---|---|---|
| **Space Group** | *P-3m1*; (#164) |  | *Pmnm*; (#59) |  | *P6$_3$/mcm*; (#193) |  |
| **Temp (K)** | 120(2) |  | 100 |  | 326 |  |
| **Bond Length (Å)** | Ti1-I1 (x6) | 2.8224(8) | Ti1-I1 (x2) | 2.7433(6) | Ti1-I1 (x6) | 2.7885(3) |
|  |  |  | Ti1-I2 (x2) | 2.8318(6) |  |  |
|  |  |  | Ti1-I3 (x1) | 2.7420(8) |  |  |
|  |  |  | Ti1-I4 (x1) | 2.8346(8) |  |  |
| **Bond Angle (°)** | I1-Ti1-I1 (x6) | 90.58(3) | I1-Ti1-I1 (x1) | 93.83(3) | I1-Ti1-I1 (x6) | 90.630(7) |
|  | I1-Ti1-I1 (x6) | 89.42(3) | I1-Ti1-I2 (x2) | 90.025(11) | I1-Ti1-I1 (x6) | 89.370(7) |
|  | I1-Ti1-I1 (x3) | 180.0 | I1-Ti1-I3 (x2) | 93.48(2) | I1-Ti1-I1 (x3) | 180.0 |
|  |  |  | I1-Ti1-I4 (x2) | 90.405(19) |  |  |
|  |  |  | I2-Ti1-I2 (x1) | 85.82(3) |  |  |
|  |  |  | I2-Ti1-I3 (x2) | 90.545(18) |  |  |
|  |  |  | I2-Ti1-I4 (x2) | 85.291(17) |  |  |
|  |  |  | I1-Ti1-I2 (x2) | 174.24(3) |  |  |
|  |  |  | I3-Ti1-I4 (x1) | 174.31(3) |  |  |

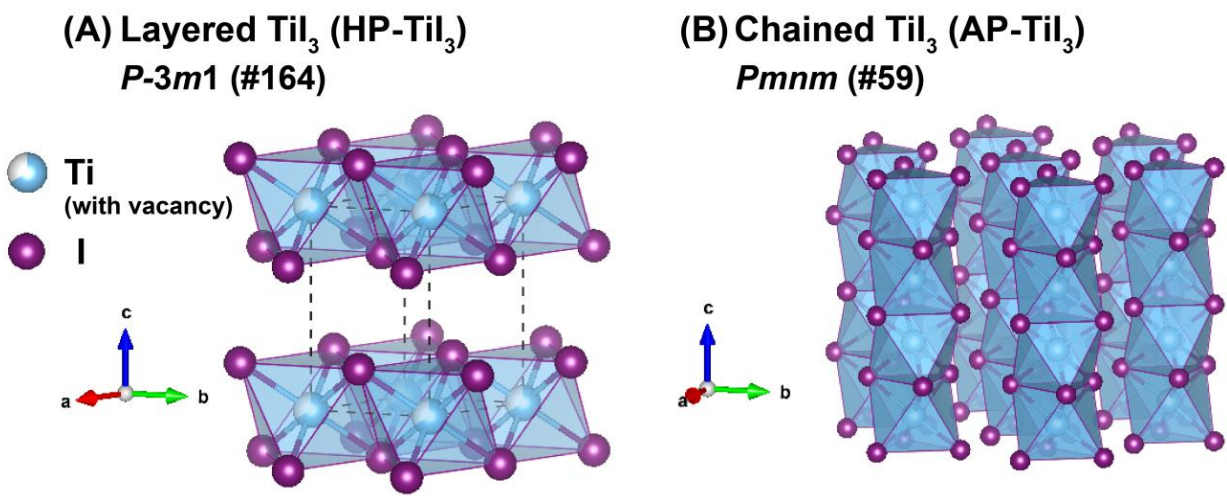

**Figure 1. The crystal structures of TiI₃.** **(A)** The average *P*-3*m*1 unit cell of HP-TiI3, refined from SCXRD measurements; **(B)** The *Pmnm* unit cell of the 1D-chain structure of AP-TiI3 [22]. Titanium is represented by the light blue spheres and iodine by the purple spheres, while vacancies are shown by white color.

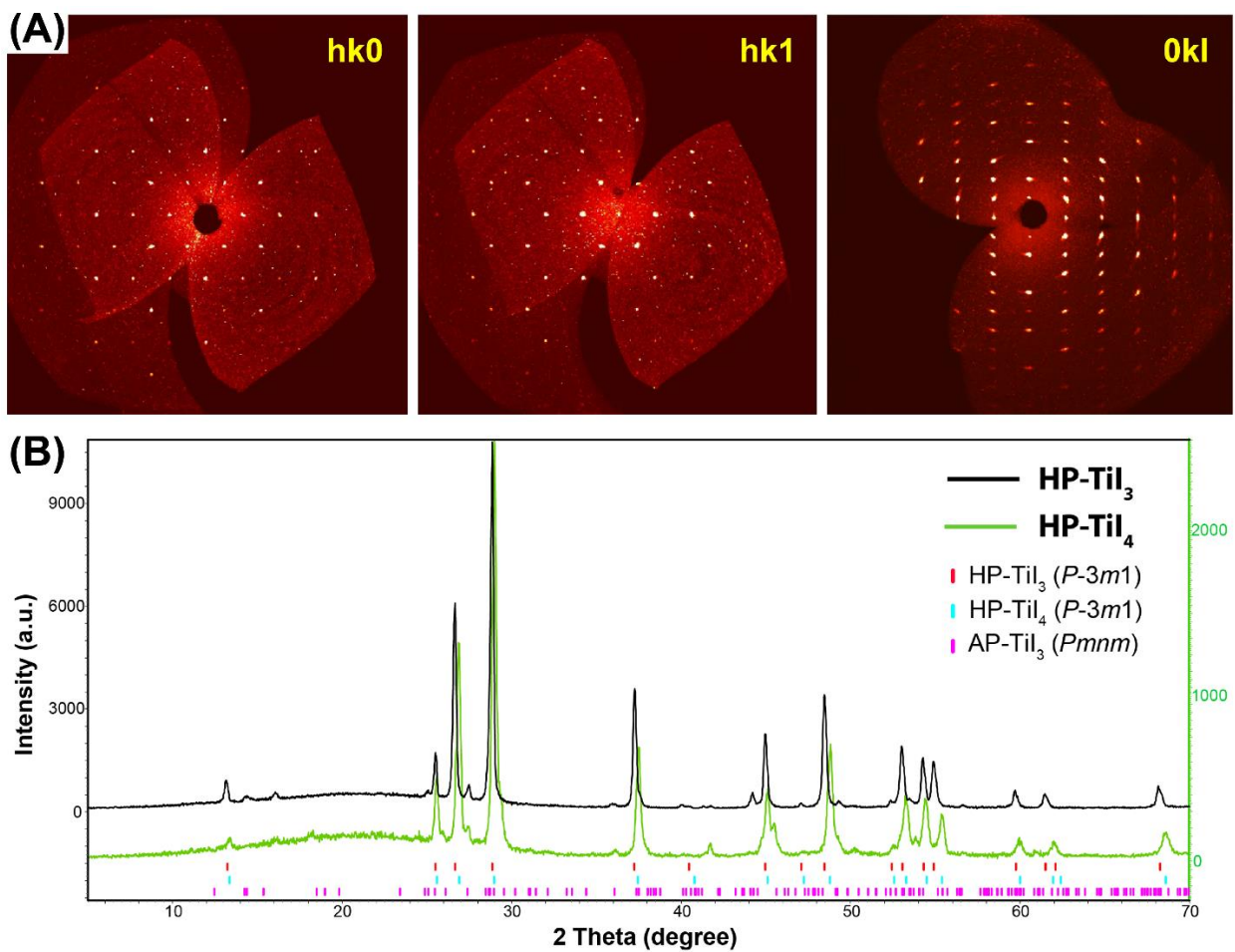

**Figure 2. The diffraction characterization of HP-TiI₃ and HP-TiI₄. (A).** The hk0 (left) hk1 (middle) and 0kl reciprocal lattice planes of single crystal HP-TiI₃ at 120 K. **(B)** The laboratory PXRD patterns of HP-TiI₃ (black pattern with left *y*-axis) and HP-TiI₄ (green pattern with right *y*-axis), with the calculated Bragg reflection tics of HP-TiI₃, HP-TiI₄, and AP-TiI₃ as a small amount of impurity.

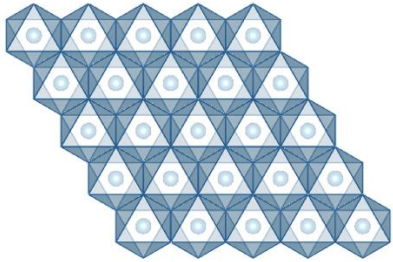 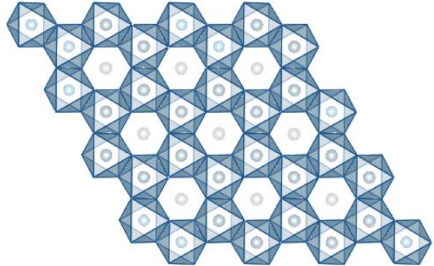 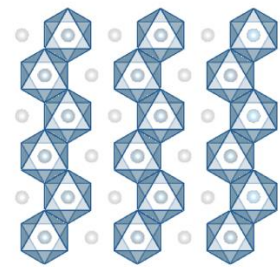

**Figure 3. Three different arrangements of the potential in-plane lattices. (A)** A fully occupied triangular lattice ($MX_2$) of the $CdI_2$ type; **(B)** A 2/3 occupied honeycomb lattice ($MX_3$) as is observed for $BiI_3$, and **(C)** A 1/2 occupied in-plane zig-zag chain lattice as is found for $TiI_4$. [3,14,24]

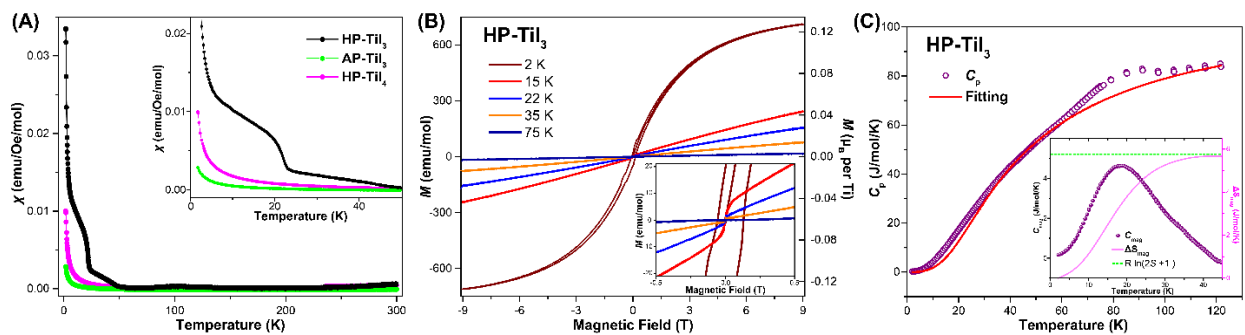

**Figure 4. The magnetic and thermodynamic characterization of TiI₃. (A)** The temperature dependent magnetic susceptibility measured with zero-field cooling from 1.8 to 300 K under 1 kOe of HP-TiI₃ (black), AP-TiI₃ (green), and HP-TiI₄ (pink). 1.8 to 50 K range is presented in the inset to exhibit the magnetic features; **(B)** Magnetic moments measured versus magnetic field from -9 T to 9 T under different temperatures, on polycrystalline HP-TiI₃ powders, with μ$_B$ per Ti unit shown on the right axis. Zoomed-in view of -0.5 T to 0.5 T range is shown in the inset. **(C)** Heat capacity of HP-TiI₃ measured under zero field from 2 to 120 K. Red curve represents the Debye model fitting of phonon contribution, with Eq (1). The subtracted C$_{mag}$ curve is plotted in the inset versus temperature from 2 to 45 K. Corresponding entropy change is exhibited in pink curve, with a comparison the value of Heisenberg limit Rln(2$S$+1) = Rln(2) as the green dashed line.